# Realization of Flattened Structural Luneburg Lens Based on Quasi-Conformal Transformation

Liuxian Zhao[a], Miao Yu[a,b,*]

[a] Institute for Systems Research, University of Maryland, College Park, MD, 20742, USA

[b] Department of Mechanical Engineering, University of Maryland, College Park, Maryland 20742, USA

*Author to whom correspondence should be addressed: mmyu@umd.edu

Conventional structural Luneburg lens is a symmetric circular gradient-index lens with refractive indices decreasing from the centre along the radial direction. In this paper, a flattened structural Luneburg lens (FSLL) based on structural thickness variations is designed by using the quasi-conformal transformation (QCT) technique. Through numerical simulations and experimental studies, the FSLL is demonstrated to have excellent beam steering performance for the manipulation of flexural wave propagation at desired angles.

Structural beamforming is a critical technique in the areas of structural health monitoring and non-destructive testing, which has attracted widespread interests [1, 2]. To obtain multi-directional beamforming, conventional phase array techniques based on phase-shift systems are employed, which often require complicated design, expensive fabrication, and high input power [3, 4]. With the development of graded-index metamaterials in recent years [5-11], Luneburg lenses are receiving considerable attentions for the manipulation of not only optical waves [12, 13] and acoustic waves [14-16], but also structural waves [17-19]. Compared with the conventional phase array techniques, structural Luneburg lens, as an attractive graded-index lens, can be used to achieve arbitrary directional beamforming without the need for any phase-shift systems [20, 21].

The conventional structural Luneburg lens is a circular gradient-index lens and every point on the lens surface can act as a source location for generating plane wave propagation on the opposite side of the lens. However, the applications of the conventional structural Luneburg lens are limited due to its circular aperture. The lens can hardly accommodate flat feeding sources used in practical applications. Furthermore, as the lens is often an integral part of the test structure, the circular aperture requires the placement of transducers inside the structure, which is not always feasible.

In this study, we propose a flattened structural Luneburg lens (FSLL) for flexural wave beamforming. The FSLL is designed by using the Quasi-Conformal Transformation (QCT) technique, which allows to modify the structure geometry and refractive index distribution with minimum anisotropy and reduced computational complexity [22-25]. The QCT technique has been explored in the design of optical and acoustic metamaterial structures [26-31]. The design principle of our FSLL based on the QCT is illustrated in Figure 1. The original circular structural Luneburg lens, denoted as the virtual space in Figure 1(a), is transformed into an FSLL denoted as the physical space in Figure 1(b). Feeding sources (e.g., a line source, several individual sources, or a moving single source) can be attached to the flat side of the lens to achieve beamforming of different angles. The refractive index distribution of the FSLL in the physical space is calculated by solving Laplace's equation with Dirichlet and Neumann boundary conditions as given below [32]:

$$\begin{cases} AB|_x = EF|_x = x' \\ AF|_y = BC|_y = CD|_y = DE|_y = y' \\ \vec{n} \cdot \nabla_x|_{BC,CD,DE,AF} = 0 \\ \vec{n} \cdot \nabla_y|_{AB,EF} = 0 \end{cases}, \quad (1)$$

where (x, y) and (x', y') are the coordinates of the virtual space and the physical space, respectively, $\vec{n}$ represents the outward normal vector to the surface boundaries, and $\nabla$ is the gradient operator.

We then obtain the effective refractive index $n'$ of the FSLL as the following [33, 34]:

$$n'^2 = \frac{n^2}{\det(J) n_0^2}, \quad (2)$$

where $n$ and $n_0$ represent the refractive indices of the circular structural Luneburg lens and the background medium, respectively, and $J$ is the Jacobian matrix.

A commercial software COMSOL is used to solve the Laplace equation with the corresponding boundary conditions to obtain the refractive index distribution of the FSLL. The normalized refractive index distribution ($n/n_0$) of the virtual space is shown in Figure 1(c). The transformed shape and normalized refractive index distribution ($n'/n_0$) of the physical space are shown in Figure 1(d).

The FSLL is constructed by using a variable thickness structure defined in a thin plate. Based on the obtained refractive index distribution of the FSLL in Figure 1(d), the range of the refractive index is between 1 and 1.82. This refractive index variation can be realized by varying the structural thickness. The same approach has been employed for achieving Acoustic

Black Holes [35-39] and other types of structural lenses [21, 40]. By using Eq. (2), a graded variation of thickness $h$ can be obtained as [21]:

$$h = \frac{h_0}{n'^2} ,\qquad(3)$$

where $h_0$ is the constant plate thickness.

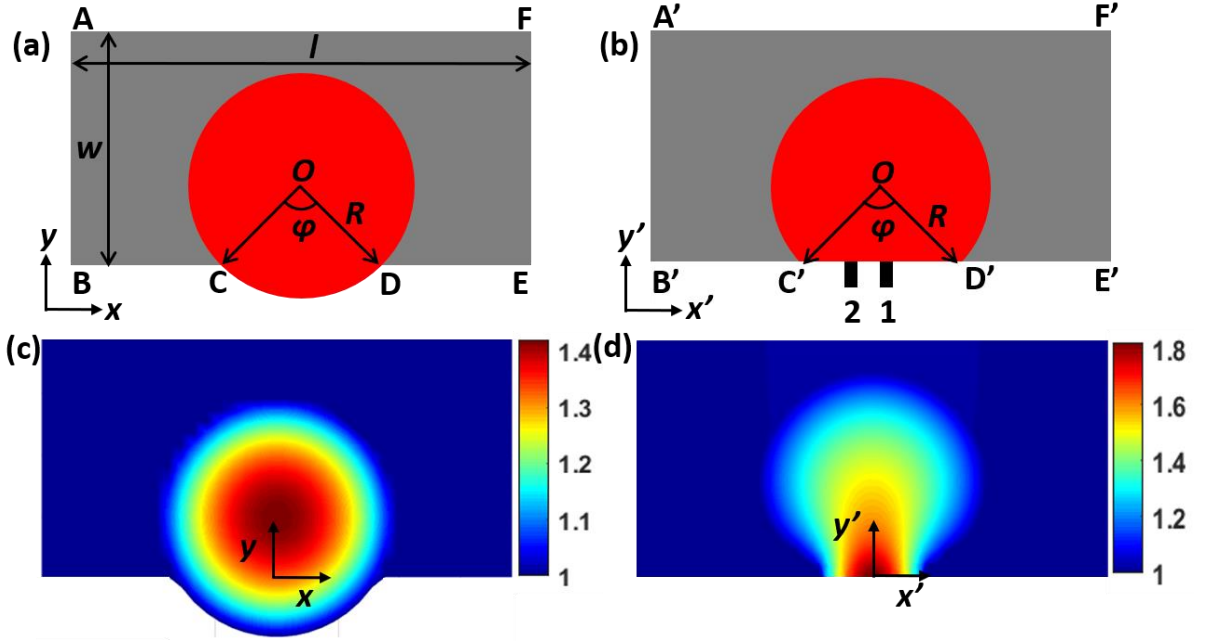

**Figure 1: Design of flattened structural Luneburg lens based on QCT. Schematic of the structural Luneburg lens designs in (a) virtual and (b) physical spaces. The red areas indicate the original circular and the flattened structural Luneburg lenses. Refractive index distributions for (c) circular structural Luneburg lens and (d) flattened structural Luneburg lens. $R$ is the radius of the lens, $\varphi$ is the open angle, and dimension of the entire space is $l \times w = 4R \times 2R$. The results in (c) and (d) were obtained based on Radius $R = 5$ cm and open angle $\varphi = 90°$.**

Note that our proposed FSLL is a broadband device in principle. There is no upper bound frequency for the FSLL due to its continuously varying refractive indices. On the other hand, the lower bound frequency is limited by the scattering regime, which requires the wavelength $\lambda$ to be less than the radius of the lens $R$.

For proof-of-concept, the following parameters are chosen in the following studies: radius $R = 5$ cm, open angle $\varphi = 90°$, constant thin plate thickness $h_0 = 0.004$ m, and dimension of the entire plate $l_0 \times w_0 \times h_0 = 0.45$ m $\times$ 0.3 m $\times$ 0.004 m. The open angle defines the azimuthal angle range of the beamforming, which can be designed to be any value from 0° to 180°. Here, for the designed open angle of 90°, the beam can be steered over the azimuthal angle range from -45° to 45°, when a feeding source moves along the flat side of the lens. For simplicity without losing the generality, two feeding source locations were considered, as shown in Figure 1(b). They are labelled as 1 ($x'=0$ mm, $y'=0$ mm) and 2 ($x'=-8$ mm, $y'=0$ mm), which correspond to the steering angles of approximately 0° and 16.5°, respectively.

Full 3D wave simulations were conducted by using commercial software COMSOL. Both frequency and time domain analyses were carried out to show the flexural wave beamforming by using the FSLL. Perfectly Matched Layers (PML) and Low Reflecting Boundaries (LRB) were used to reduce the boundary reflections in frequency response and time domain analysis correspondingly. The frequency domain analysis was performed for the frequency range from 20 - 80 kHz. Based on the lens radius $R = 5$ cm, we chose the lower bound frequency to be 20 kHz. Although there is no theoretical upper bound frequency, due to the limitation of the numerical simulations, which require 5-10 elements for simulating the smallest wavelength, an upper bound frequency of 80 kHz was chosen in this study.

The obtained waveforms at the frequency of 40 kHz were shown in Figures 2 (a) and (b). It can be seen that the main wave energy propagates along the steering angles of 0° and 16.5° (the black dot dashed rectangles), respectively, when point sources 1 and 2 are excited. The phase plots are presented in Figures 2 (c) and (d), which clearly show that the circular phase originated from the source is transformed into flat phases along the steering angle directions (the black dot dashed rectangles). These results demonstrate that the outgoing waves are collimated at the steering angles of 0º and 16.5º. In order to characterize the broadband beamforming performance of the FSLL, the phase data along the direction that is perpendicular to the beamforming direction (e.g., white dashed lines in Figure 2s (c) and (d)) are plotted in Figures 2 (e) and (f) over a broad frequency range of 20 -80 kHz. These results clearly show that the phase data obtained for the outgoing beams along the corresponding steering angle directions are constant over the entire simulated frequency range, demonstrating the broadband beamforming characteristic of the FSLL.

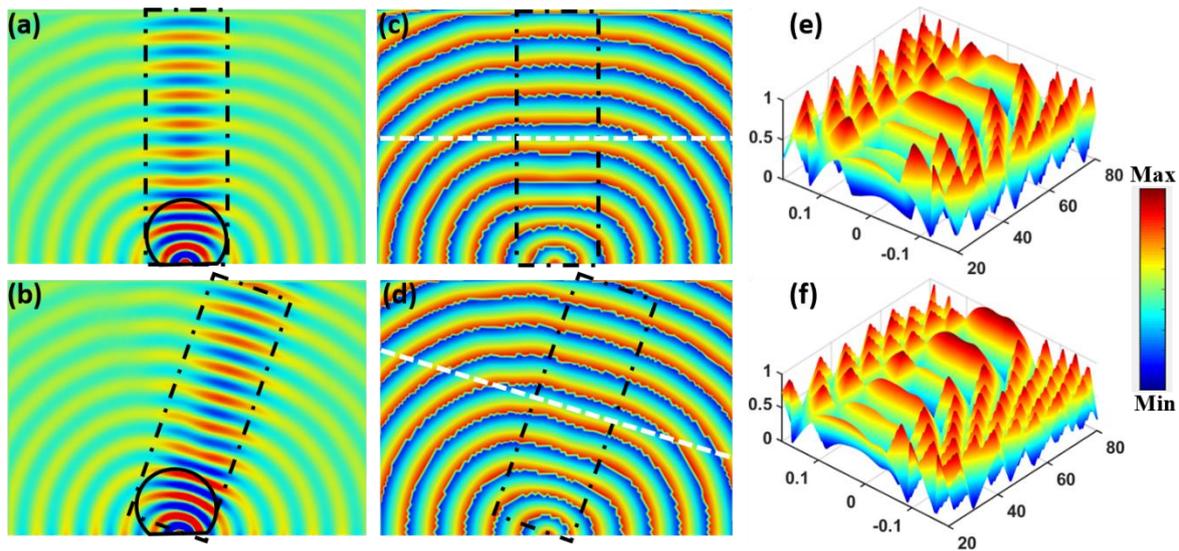

**Figure 2: Frequency domain analysis of flexural wave beamforming at steering angles of 0º and 16.5º by using the FSLL. Waveforms obtained for beam steering angles of (a) 0º and (b) 16.5º at 40 kHz. Phase distributions obtained for beam steering angles of (c) 0º and (d) 16.5º at 40 kHz. Phase data over the frequency range of 20 - 80 kHz obtained for the steering angles of (e) 0º and (f) 16.5º. The dot dashed black rectangles in (a)-(d) indicate the areas of beamforming, and the dashed white lines in (c) and (d) are perpendicular to the beamforming directions.**

Furthermore, time domain analysis was performed to examine the flexural wave beam steering at different time steps. A signal of 3-count tone bursts at 40 kHz was used as the excitation source. The full field waveforms of propagation at different time steps obtained for the steering angles of 0º and 16.5º are shown in Figures 3 (a)-(c) and Figures 3 (d)-(f), respectively. Again, the main wave energy is shown to propagate along the directions of the steering angles (0º and 16.5º). From $t = 0$ ms to 0.05 ms, the circular flexural waves originated from the point source propagate forward. At $t = 0.1$ ms, the flexural waves interact with the FSLL and the waveforms gradually become collimated. After $t = 0.15$ ms, the waves propagate forward as plane waves.

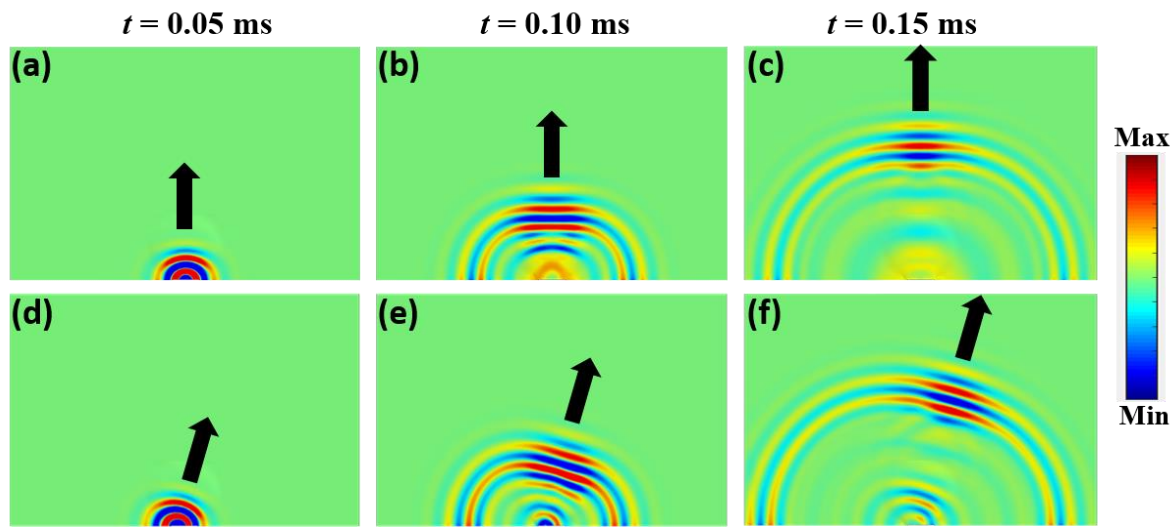

**Figure 3: Transient analysis of flexural wave beamforming at 40 kHz by using the FSLL. (a) –(c) Responses obtained for steering angle of 0º at time instants of $t = 0.05$ ms, 0.10 ms, and 0.15 ms. (d) –(f) Responses obtained for steering angle of 16.5º at time instants of $t = 0.05$ ms, 0.10 ms, and 0.15 ms.**

Experimental studies were also carried out to characterize the performance of an FSLL fabricated on a thin plate (6061 aluminium from McMaster-Carr) with dimensions of 0.6 m × 0.3 m× 0.004 m. The experimental setup and the fabricated FSLL are shown in Figure 4. The four sides of the plate were covered with plastilina modelling clay to minimize boundary reflections. Two circular piezoelectric discs (12 mm in diameter and 0.6 mm in thickness from STEMiNC Corp.) were used as point sources, which were placed at the two feeding source locations (1 ($x'$=0 mm, $y'$=0 mm) and 2 ($x'$=-8 mm, $y'$=0 mm)) shown in Figure 1(b). A piezoelectric transducer was bonded to the plate by using an adhesive (2P-10 adhesive from Fastcap, LLC). A scanning laser Doppler vibrometer (SLDV) (PSV-400 from Polytec) was used to measure the propagating wave field by recording the out-of-plane particle velocity on the plate as shown in Figure 4(a). The scanning was performed on the front surface (flat side) of the plate, as shown in Figure 4(c). The back surface of the plate with the FSLL is shown in Figure 4(b). In the experiment, a voltage signal of 3-count tone bursts was used to excite the transducer at 40 kHz. Transient responses were obtained to examine the beamforming performance of the FSLL. Note that only one frequency ($f$ = 40 kHz) was used in the experiment. For higher frequencies, the strong reflections from the plate boundary can obfuscate the performance of beamforming. On the other hand, the experimental studies at lower frequencies would require a much larger plate, which causes unnecessary complication for the experiment. Therefore, the frequency of 40 kHz was chosen as a good compromise.

The measured waveforms of the out-of-plane particle velocity fields at different time instants for two steering angles of 0º and 16.5º are shown in Figures 4 (d)-(f) and Figures 4(g)-(i), respectively. These experimental results are in excellent agreement with the numerical simulation results shown in Figure 3, which further validate the beam steering capability of the FSLL.

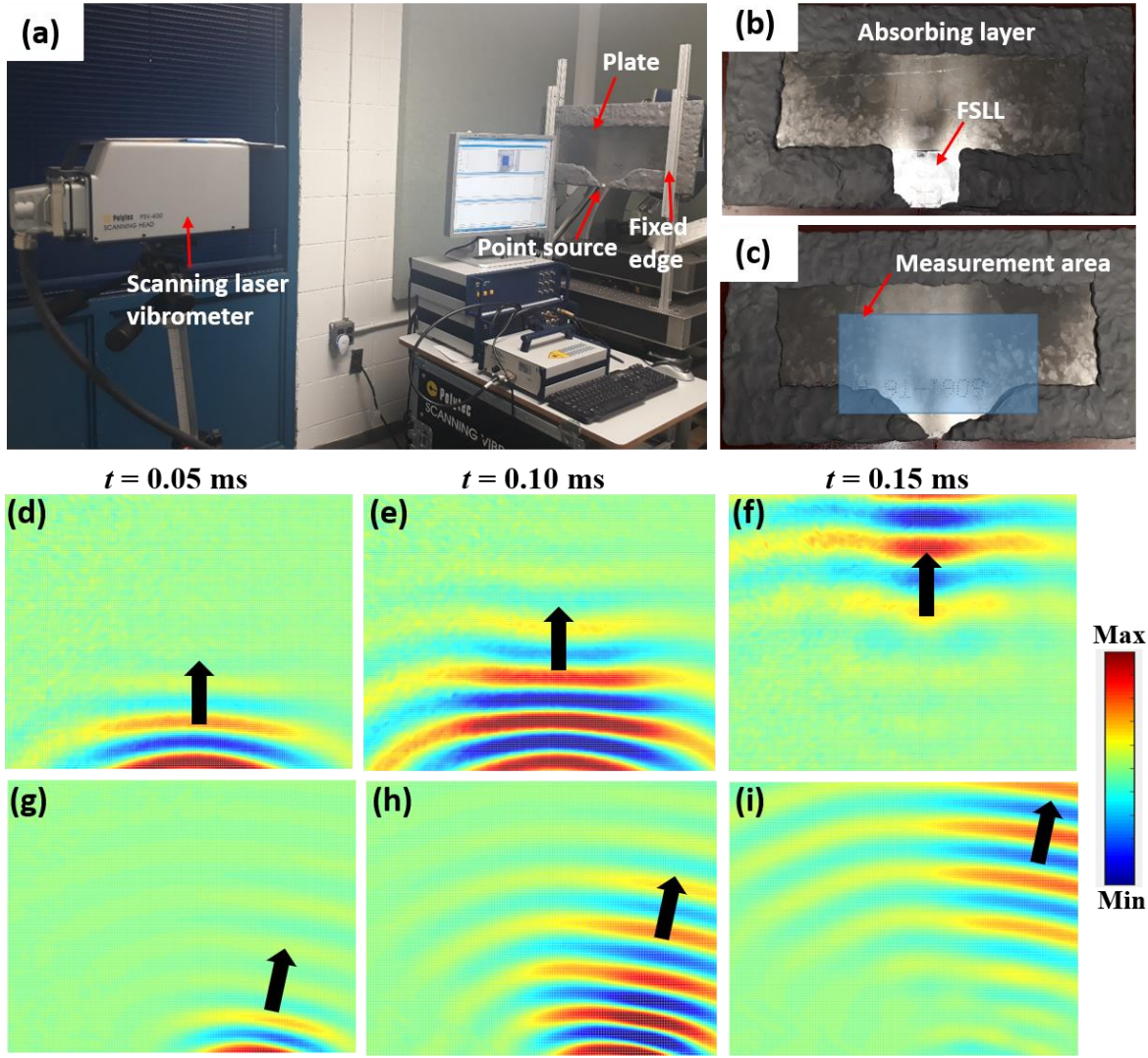

**Figure 4: Experimental studies of beamforming using the FSLL. (a) Photograph of the experimental setup. (b) Fabricated FSLL on the plate (back surface). (c) Front surface of the plate with the measurement area indicated in blue. (d)-(f) Transient responses obtained for steering angle of 0º at time instants of $t$ = 0.05 ms, 0.10 ms, and 0.15 ms. (g)-(i) Transient responses obtained for steering angle of 16.5º at time instants of $t$ = 0.05 ms, 0.10 ms, and 0.15 ms.**

We numerically and experimentally demonstrated a flattened structural Luneburg lens (FSLL) for beam steering of flexural waves. The FSLL was designed by using the quasi-conformal transformation (QCT) technique with Dirichlet and Neumann boundary conditions.

The flattened lens surface allows easy accommodation of feeding sources for practical applications. In this study, the FSLL was fabricated by using a variable thickness structure defined in a thin plate, which rendered a continuous change of the refractive index for smooth manipulation of flexural wave propagation. The numerical simulation results indicate that the FSLL can successfully perform beamforming at the designed steering angles over a broad frequency range. The experimental results further validated the beam steering performance of the FSLL at a particular frequency, which were in good agreement with the numerical simulation results. The FSLL based beamforming could benefit many potential applications including ultrasonic imaging, structural health monitoring, and non-destructive testing.

**Conflict of Interest**

The authors declare no conflict of interest.

**Reference**

1. Medda, A. and V. DeBrunner, *Near-field Sub-band Beamforming for Damage Detection in Bridges.* Structural Health Monitoring, 2009. **8**(4): p. 313-329.
2. Poozesh, P., et al., *Structural health monitoring of wind turbine blades using acoustic microphone array.* Structural Health Monitoring, 2016. **16**(4): p. 471-485.
3. Rocha, B., et al., *Design and Development of a Phased Array System for Damage Detection in Structures*, in *Structural Health Monitoring for Advanced Composite Structures*. 2017, WORLD SCIENTIFIC (EUROPE). p. 153-189.
4. Huan, Q., et al., *A high-resolution structural health monitoring system based on SH wave piezoelectric transducers phased array.* Ultrasonics, 2019. **97**: p. 29-37.
5. Chen, Y., et al., *Enhanced acoustic sensing through wave compression and pressure amplification in anisotropic metamaterials.* Nature Communications, 2014. **5**(1): p. 5247.
6. Peng, S., et al., *Acoustic far-field focusing effect for two-dimensional graded negative refractive-index sonic crystals.* Applied Physics Letters, 2010. **96**(26): p. 263502.
7. Liang, Y.-J., et al., *An acoustic absorber implemented by graded index phononic crystals.* Journal of Applied Physics, 2014. **115**(24): p. 244513.
8. Jin, Y., et al., *Flat acoustics with soft gradient-index metasurfaces.* Nature Communications, 2019. **10**(1): p. 143.
9. Wu, L.-Y. and L.-W. Chen, *An acoustic bending waveguide designed by graded sonic crystals.* Journal of Applied Physics, 2011. **110**(11): p. 114507.
10. Lee, D., et al., *Retrieving continuously varying effective properties of non-resonant acoustic metamaterials.* Applied Physics Express, 2019. **12**(5): p. 052008.

11. Zhao, L. and S. Zhou, *Compact Acoustic Rainbow Trapping in a Bioinspired Spiral Array of Graded Locally Resonant Metamaterials.* Sensors, 2019. **19**(4).
12. Loo, Y.L., et al., *Broadband microwave Luneburg lens made of gradient index metamaterials.* Journal of the Optical Society of America A, 2012. **29**(4): p. 426-430.
13. Liang, M., et al., *A 3-D Luneburg Lens Antenna Fabricated by Polymer Jetting Rapid Prototyping.* IEEE Transactions on Antennas and Propagation, 2014. **62**(4): p. 1799-1807.
14. Wang, C.-F., et al., *Wideband Acoustic Luneburg Lens Based on Graded Index Phononic Crystal.* 2015(57564): p. V013T16A023.
15. Park, C.M. and S.H. Lee, *Acoustic Luneburg lens using orifice-type metamaterial unit cells.* Applied Physics Letters, 2018. **112**(7): p. 074101.
16. Fu, Y., et al., *Compact acoustic retroreflector based on a mirrored Luneburg lens.* Physical Review Materials, 2018. **2**(10): p. 105202.
17. Torrent, D., Y. Pennec, and B. Djafari-Rouhani, *Omnidirectional refractive devices for flexural waves based on graded phononic crystals.* Journal of Applied Physics, 2014. **116**(22): p. 224902.
18. Zhao, L., C. Lai, and M. Yu, *Modified structural Luneburg lens for broadband focusing and collimation.* Mechanical Systems and Signal Processing, 2020. **144**: p. 106868.
19. Zhao, L. and M. Yu, *Bi-functional Structural Luneburg Lens for Broadband Structural Wave Cloaking and Waveguide*. 2020: arXiv preprint arXiv:2003.10632.
20. Tol, S., F.L. Degertekin, and A. Erturk, *Phononic crystal Luneburg lens for omnidirectional elastic wave focusing and energy harvesting.* Applied Physics Letters, 2017. **111**(1): p. 013503.
21. Climente, A., D. Torrent, and J. Sánchez-Dehesa, *Gradient index lenses for flexural waves based on thickness variations.* Applied Physics Letters, 2014. **105**(6): p. 064101.
22. Landy, N.I., N. Kundtz, and D.R. Smith, *Designing Three-Dimensional Transformation Optical Media Using Quasiconformal Coordinate Transformations.* Physical Review Letters, 2010. **105**(19): p. 193902.
23. Junqueira, M.A.F.C., et al., *Three-dimensional quasi-conformal transformation optics through numerical optimization.* Optics Express, 2016. **24**(15): p. 16465-16470.
24. Kwon, D., *Quasi-Conformal Transformation Optics Lenses for Conformal Arrays.* IEEE Antennas and Wireless Propagation Letters, 2012. **11**: p. 1125-1128.
25. Mei, Z.L., J. Bai, and T.J. Cui, *Experimental verification of a broadband planar focusing antenna based on transformation optics.* New Journal of Physics, 2011. **13**(6): p. 063028.
26. Biswas, S., et al., *Realization of modified Luneburg lens antenna using quasi-conformal transformation optics and additive manufacturing.* Microwave and Optical Technology Letters, 2019. **61**(4): p. 1022-1029.
27. Li, Y. and Q. Zhu, *Luneburg lens with extended flat focal surface for electronic scan applications.* Optics Express, 2016. **24**(7): p. 7201-7211.
28. Wan, X., et al., *A broadband transformation-optics metasurface lens.* Applied Physics Letters, 2014. **104**(15): p. 151601.
29. Silva, D.G., et al., *Full three-dimensional isotropic carpet cloak designed by quasi-conformal transformation optics.* Optics Express, 2017. **25**(19): p. 23517-23522.
30. Li, J. and J.B. Pendry, *Hiding under the Carpet: A New Strategy for Cloaking.* Physical Review Letters, 2008. **101**(20): p. 203901.
31. Dong, H.Y., et al., *Realization of broadband acoustic metamaterial lens with quasi-conformal mapping.* Applied Physics Express, 2017. **10**(8): p. 087202.
32. Zhao, L., et al., *Ultrasound beam steering with flattened acoustic metamaterial Luneburg lens.* Applied Physics Letters, 2020. **116**(7): p. 071902.
33. Dong, H.Y., et al., *A broadband planar acoustic metamaterial lens.* Physics Letters A, 2019. **383**(16): p. 1955-1959.
34. Sun, Z., et al., *Quasi-isotropic underwater acoustic carpet cloak based on latticed pentamode metafluid.* Applied Physics Letters, 2019. **114**(9): p. 094101.

35. Zhang, Y., et al., *Ultralight phononic beam with broad low-frequency band gap using complex lattice of acoustic black holes.* Applied Physics Express, 2019. **12**.
36. Tang, L. and L. Cheng, *Periodic plates with tunneled Acoustic-Black-Holes for directional band gap generation.* Mechanical Systems and Signal Processing, 2019. **133**: p. 106257.
37. Zhao, L., F. Semperlotti, and S.C. Conlon. *Enhanced vibration based energy harvesting using embedded acoustic black holes*. in *Proc.SPIE*. 2014.
38. Zhao, L., S.C. Conlon, and F. Semperlotti, *Broadband energy harvesting using acoustic black hole structural tailoring.* Smart Materials and Structures, 2014. **23**(6): p. 065021.
39. Zhao, L., *Low-frequency vibration reduction using a sandwich plate with periodically embedded acoustic black holes.* Journal of Sound and Vibration, 2019. **441**: p. 165-171.
40. Zareei, A., et al., *Continuous profile flexural GRIN lens: Focusing and harvesting flexural waves.* Applied Physics Letters, 2018. **112**(2): p. 023901.